\documentclass[preprint]{elsarticle}
\usepackage{xcolor}
\usepackage{enumitem}
\usepackage{graphicx}
\usepackage{tabularx}
\usepackage{placeins}
\usepackage[load=addn, separate-uncertainty=true]{siunitx}
\usepackage[numbers]{natbib}
\usepackage{amsmath}
\usepackage{nicefrac}
\usepackage{booktabs}
\usepackage{upgreek}
\usepackage[skip=0.25\baselineskip]{caption}

\usepackage{lineno}
\usepackage[hidelinks]{hyperref}
\modulolinenumbers[5]

\newcommand*\dif{\mathop{}\!\mathrm{d}}
\DeclareSIUnit\liter{\ensuremath{\ell}}
\DeclareSIUnit\litre{\ensuremath{\ell}}

\newcommand*\estimates{\mathrel{\hat{=}}}
\newcommand*\estimatesapprox{\mathrel{\hat{\approx}}}

\journal{Vacuum}

\begin{document}

\setlist[description]{font=\normalfont\itshape\textbullet\space}

\begin{frontmatter}

\title{Neutral tritium gas reduction in the KATRIN differential pumping sections%
%\tnoteref{mytitlenote}%
}
%\tnotetext[mytitlenote]{Technical Paper, suggestion: KATRIN publication category B.}


\author[mymainaddress]{A.~Marsteller}\corref{mycorrespondingauthor}
\cortext[mycorrespondingauthor]{Corresponding author}
\ead{alexander.marsteller@kit.edu}

\author[mymainaddress]{B.~Bornschein}
\author[mytertiaryaddress]{L.~Bornschein}
\author[mysecondaryaddress]{G.~Drexlin}
\author[mytertiaryaddress]{F.~Friedel}
\author[itepaddress]{R.~Gehring}
%\author[mytertiaryaddress]{W.~Gil}
\author[ittkaddress]{S.~Grohmann}
\author[mytertiaryaddress]{R.~Gumbsheimer}
\author[hackenjosaddress]{M.~Hackenjos}
\author[mytertiaryaddress]{A.~Jansen}
\author[kosmideraddress]{A.~Kosmider}
\author[mysecondaryaddress]{L.~LaCascio}
\author[mytertiaryaddress]{S.~Lichter}
\author[mytertiaryaddress]{K.~M\"uller}
\author[mymainaddress]{F.~Priester}
\author[mytertiaryaddress]{R.~Rinderspacher}
\author[mymainaddress]{M.~R\"ollig}
\author[mymainaddress]{C.~R\"ottele}
\author[sharipovaddress]{F.~Sharipov}
\author[mymainaddress]{M.~Sturm}
\author[mymainaddress]{S.~Welte}
\author[mysecondaryaddress]{J.~Wolf}

\address[mymainaddress]{Institute for Nuclear Physics, Tritium Laboratory Karlsruhe(IKP-TLK), Karlsruhe Institute of Technology (KIT), Hermann-von-Helmholtz-Platz 1, 76344 Eggenstein-Leopoldshafen, Germany}
\address[mytertiaryaddress]{Institute for Nuclear Physics (IKP), Karlsruhe Institute of Technology (KIT), Hermann-von-Helmholtz-Platz 1, 76344 Eggenstein-Leopoldshafen, Germany}
\address[mysecondaryaddress]{Institute of Experimental Particle Physics (ETP), Karlsruhe Institute of Technology (KIT), Wolfgang-Gaede-Str. 1, 76131 Karlsruhe, Germany}
\address[itepaddress]{Institute for Technical Physics (ITEP), Karlsruhe Institute of Technology (KIT), Hermann-von-Helmholtz-Platz 1, 76344 Eggenstein-Leopoldshafen, Germany}
\address[ittkaddress]{Institute of Technical Thermodynamics and Refrigeration (ITTK), Karlsruhe Institute of Technology (KIT), Engler-Bunte-Ring 21, 76131 Karlsruhe, Germany}
\address[sharipovaddress]{Departamento de F\'isica, Universidade Federal do Paran\'a, Caixa Postal 19044, 81531-980, Curitiba, Brazil}
\address[kosmideraddress]{Helmholtz Association Berlin Office, Anna-Louisa-Karsch-Str. 2 10178 Berlin, formerly Institute for Nuclear Physics (IKP), Karlsruhe Institute of Technology (KIT)}
\address[hackenjosaddress]{No longer affiliated with KIT, formerly Institute for Nuclear Physics (IKP), Karlsruhe Institute of Technology (KIT), Hermann-von-Helmholtz-Platz 1, 76344 Eggenstein-Leopoldshafen, Germany}

\begin{abstract}
The KArlsruhe TRItium Neutrino experiment (KATRIN) aims to measure the effective electron anti-neutrino mass with an unprecedented sensitivity of {\SI[allow-number-unit-breaks]{0.2}{e\volt\per c\squared}}, using $\upbeta$-electrons from tritium decay. 
The electrons are guided magnetically by a system of superconducting magnets through a vacuum beamline from the windowless gaseous tritium source through differential and cryogenic pumping sections to a high resolution spectrometer and a segmented silicon pin detector. 
At the same time tritium gas has to be prevented from entering the spectrometer. 
Therefore, the pumping sections have to reduce the tritium flow by more than \num{14} orders of magnitude. 
This paper describes the measurement of the reduction factor of the differential pumping section performed with high purity tritium gas during the first measurement campaigns of the KATRIN experiment. 
The reduction factor results are compared with previously performed simulations, as well as the stringent requirements of the KATRIN experiment.

\end{abstract}


\end{frontmatter}

%\linenumbers

\section{Introduction}
	
	Neutrinos are the lightest and most abundant of the known massive elementary particles in the universe. 
	They played a crucial role in the evolution of large-scale structures \cite{Zeng2019} in the early universe. 
	Although, it is established since the 1990s that the different neutrino flavour species are related to individual compositions of the neutrino mass eigenstates \cite{Fukuda1998}, of which at least two are non-vanishing, the absolute values are still unknown. 
	At present, upper limits are available from cosmic surveys, such as the cosmic microwave background \cite{Valentino2016}, and from direct neutrino mass searches using $\upbeta$-decays \cite{Tanabashi2018, Aker2019}. 
	Neutrino oscillation experiments, which are only sensitive to the difference between the neutrino mass-squared, can provide lower limits \cite{Salas2018}. %formulierung angelehnt an PDG, rpp2019-rev-neutrino-mixing.pdf; paige 42, chapter 14.9
	The determination of the absolute neutrino mass value is still one of the most crucial questions in particle and astroparticle physics.

	The Karlsruhe Tritium Neutrino Experiment (KATRIN), operated at the Karlsruhe Institute of Technology (KIT), is currently the most sensitive experiment to determine the neutrino mass from the precise measurement of the kinematics of tritium-$\upbeta$-decay \cite{Aker2019} with a design sensitivity of \SI{0.2}{e\volt\per\clight\squared} \cite{KATRINCollaboration2005}. 
	A non-vanishing neutrino mass would slightly change the shape of the $\upbeta$-spectrum close the endpoint at \SI{18.6}{\kilo e\volt}.
	After the commissioning of KATRIN with deuterium measurements in autumn 2017 \cite{Arenz2018}, the first operation with low amounts of radioactive tritium started in spring 2018 \cite{Aker2020}. 
	The search for the neutrino mass with an increased tritium flow started in 2019 \cite{Aker2019}.

	\begin{figure}
	\centering
		\includegraphics[width=\textwidth]{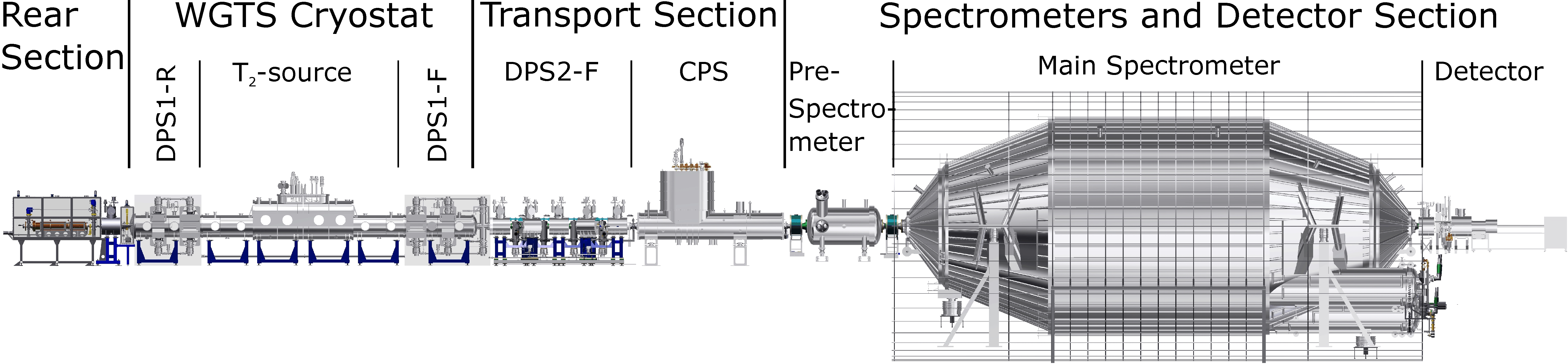}
		\caption[KATRIN beamline overview]{Overview of the KATRIN beam line.}
		\label{fig:katrin_beamline}
	\end{figure}

	The \SI{70}{\meter} long KATRIN apparatus comprises two parts (see \autoref{fig:katrin_beamline}): the tritium \textit{source and transport section} (STS) located inside the Tritium Laboratory Karlsruhe (TLK) and the tritium-free \textit{spectrometer and detector section} (SDS) located in an adjacent building \cite{Arenz2016}. 
	The STS consists of the \textit{windowless gaseous tritium source} (WGTS) cryostat, followed in downstream direction (towards SDS) by the \textit{transport and pumping section}, and upstream by the \textit{rear section}, which is part of the calibration and monitoring system (CMS). 
	Inside the WGTS beam tube tritium decays to $^3$He, isotropically emitting electrons and electron-anti-neutrinos. 
	While the neutrinos leave the beam tube without further interaction, the electrons are guided adiabatically by strong magnetic fields along the \SI{40}{\meter} long STS-beamline. Half of the electrons travel downstream to the SDS, where their energy is measured by an electrostatic spectrometer (MAC-E filter \cite{Picard1992}) with ultra-high precision (\SI{1}{e\volt} at \SI{20}{\kilo e\volt}).
	Virtually all remaining tritium gas in the beam tube is pumped out in the \textit{transport and pumping section}, before it can reach the \textit{spectrometer section}, where otherwise it would increase the background rate of the measurement. 
	Since only \num{2e-13} of all $\upbeta$-electrons have kinetic energies in the last eV below the endpoint of the spectrum, a low background rate is necessary to reach the target sensitivity. 
	Therefore, the initial tritium flow into the WGTS of \SI{1.8}{\milli\bar\litre\per\second} (here and in the following referenced to \SI{0}{\celsius}) has to be reduced by at least 14 orders of magnitude, before reaching the SDS.

\begin{figure}
			\centering
			\includegraphics[width=\textwidth]{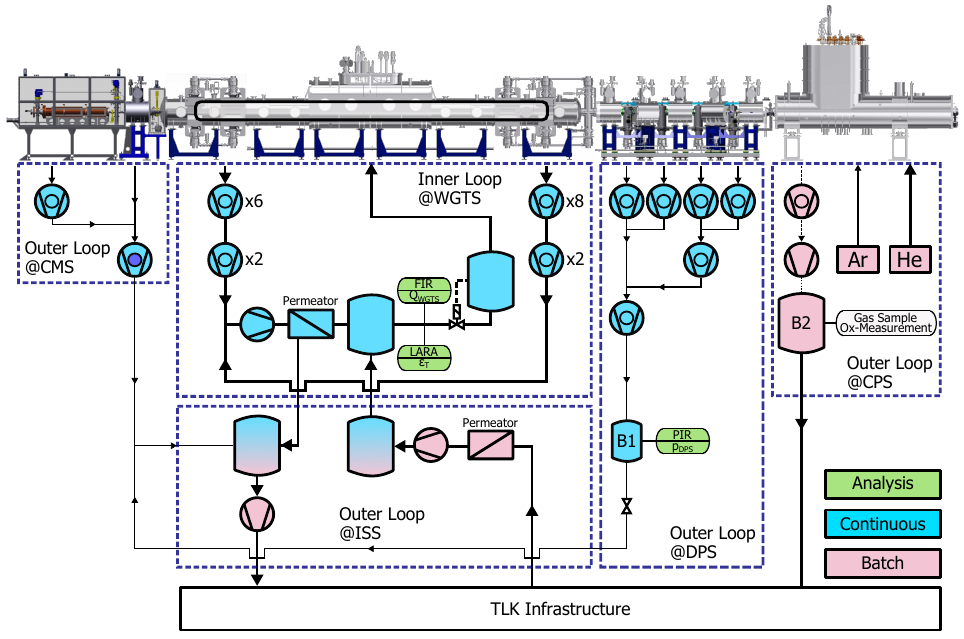}
			\caption[Simplified Flow Diagram]{Simplified flow diagram of the beamline and its pumps.
			%The TMPs at WGTS and DPS, their piping, as well as the CPS waste buffer should be shown here to be able to explain the measurements. This image should contain only very schematic depictions for the sake of clarity when explaining the measurements. \textcolor{red}{[Add flow meter and pressure sensor instrument]}
			}
			\label{fig:simplified_loops_schematic}
		\end{figure}

	An essential part of the STS is the KATRIN Loop System shown in \autoref{fig:simplified_loops_schematic}, which incorporates the pumping systems.
	It provides a closed inner loop for the ultra-pure and pressure stabilized tritium circulation through the beam tube inside the WGTS cryostat.
	Simultaneously, the outer loop serves as interlink to the tritium infrastructure of the TLK, where impure tritium gas is cleaned and stored. 
	The transport and pumping section comprises two components. First, the \textit{differential pumping section} (DPS) employs a chain of turbo molecular pumps (TMP), as shown in \autoref{fig:simplified_loops_schematic}. 
	The TMPs reduce the flow by 7 orders of magnitude. 
	Details are described in the next section.
	For lower pressures, mechanical pumping becomes inefficient. 
	Therefore, the second part is a \textit{cryogenic pumping section} (CPS), which cryosorbs the remaining tritium molecules on a \SIrange{3}{4}{\kelvin} cold argon frost layer \cite{Gil2010,Roettele2017}. 
	The Ar frost layer is regenerated regularly, before the accumulated tritium exceeds a maximum activity of \SI{3.7E10}{\becquerel} (=\SI{1}{\curie}). With the nominal tritium flow into the WGTS and the projected flow reduction by the DPS, this limit would be reached after about \SI{60}{days}. 
	A longer time between regenerations of the CPS increases the possible uptime of the KATRIN experiment. Therefore, an accurate assessment of the actual flow reduction by the DPS is important.

	This paper is focused on the reduction of the neutral tritium flow rate along the beamline by the differential pumping section, between the inlet into the beam tube at the center of the WGTS to the entrance of the CPS. 
	The overall performance of the differential pumping system of KATRIN was checked during commissioning with deuterium gas, before admitting tritium into the system. 
	The result has been confirmed during the first tritium measurement in 2018 with \SI{1}{\percent} DT in deuterium \cite{Aker2020}.
	However, these measurements allow only a rough estimate of the actual tritium reduction efficiency, due to different effective pumping speeds for DT and T$_2$. The decay rate of DT in the recovered gas after regenerating the Ar frost was used in the second measurement to determine the amount of gas accumulated in the CPS. The ratio between the accumulated gas and the integrated gas flow into the WGTS provides a measure for the reduction factor of the DPS. However, the small admixture of DT introduced a large statistical uncertainty. 
	In 2019, the first KATRIN measurements (KNM1: Mar. 27 -- May 09, 2019 \cite{Aker2019}, KNM2: Sep. 27 -- Nov. 14, 2019) with high tritium concentration (\SI{>97}{\percent}, remainder H and D) were performed, enabling the accurate determination of the DPS reduction factor for tritium.
	The results are presented in \autoref{sec:measurement} and compared to simulations \cite{Friedel2019} described in the next section.

\section{The differential pumping sections of KATRIN}
	
	\subsection{Description of the differential pumping section}
		
		%The KATRIN tritium system - called ``Inner Loop'' (IL) and ``Outer Loop'' (OL) - is a very complex piece of machinery.
		The tritium loop system is distributed along the \SI{40}{\meter} long KATRIN STS beam line and interconnects the beam line segments with each other and the TLK infrastructure \cite{Priester2015,Welte2020}. As shown in \autoref{fig:simplified_loops_schematic}, it consists of the ``Inner Loop'' (IL) and ``Outer Loop'' (OL). Components of interest for this paper are:
	
		\begin{description}
			\item[Windowless Gaseous Tritium Source (WGTS):] 
				The source of $\upbeta$-electrons in KATRIN is a gas column of tritium in a \SI{10}{\meter} long beam tube with \SI{90}{\milli\meter} diameter. In order to achieve the high statistics required for KATRIN, \SI{1.8}{\milli\bar\litre\per\second} (\SI{40}{\gram\per day}) of tritium with a purity \SI{>95}{\percent} is injected in the center of the WGTS beam tube with an inlet pressure of \SI{\approx e-3}{\milli\bar}.
				In order to minimize systematic effects, the source tube inside the WGTS cryostat is cooled down to a temperature of \SI{\approx 30}{\kelvin}. The beam tube temperature, the injection pressure, and the necessary gas throughput have to be kept stable on a level of \SI{<0.1}{\percent\per\hour}.
				%Over the \SI{16}{\meter} long WGTS cryostat, \SI{40}{\gram\per\day} tritium with a purity \SI{>95}{\percent} are circulated. The beam tube temperature has to be kept stable at \SI{\approx 30}{\kelvin} (within \SI{<0.1}{\percent}). The density profile of the free streaming gas, injected in the middle of the beam tube with \SI{\approx 1e-3}{\milli\bar} has to be kept stable to within \SI{0.1}{\percent}. 
			
			\item[Differential Pumping Section 1 (DPS1):]
				Connected on both sides to the WGTS beam tube are the first stages of the differential pumping section, the DPS1 (see \autoref{fig:simulation_model_dps_beamline}). 
				This section consists of 4 pump ports (PP) inside the WGTS cryostat (DPS1-F1, DPS1-F2, DPS1-R1, DPS1-R2) and one outside of it (PP0). A total of 14 turbomolecular pumps (TMP) of type Leybold MAG-W2800 with a pumping speed of \SI{2100}{\litre\per\second} for H$_2$ \cite{leybold_manual} are connected to these pump ports. 
				Twelve TMPs are arranged symmetrically around the WGTS beam beam tube, starting with 4 TMPs at each end, connected to DPS1-F1 and DPS-R1. The next stage includes two TMPs in DPS1-F2 and DPS1-R2, respectively. 
				The two remaining pumps are located at an additional pump port (PP0) between the WGTS cryostat and the DPS2 in downstream direction.
				The pump ports are connected via beam tube segments of \SI{1}{\meter} length and a diameter of \SI{90}{\milli\meter}.
				
				The fore-vacuum for the MAG-W2800 TMPs is provided by 4 Pfeiffer HiPace300 TMPs with a pumping speed of \SI{220}{\litre\per\second} for H$_2$ \cite{pfeiffer_manual}. These pumps are in turn pumped by cascaded fore-pumps, combining a Normetex\textsuperscript{\textregistered} scroll pump and a metal bellows pump \cite{Chang2011} (see \autoref{fig:simplified_loops_schematic}).
				Gas pumped out by this system is purified by a PdAg permeator and then re-injected into the WGTS (see \cite{Priester2020} for details). The WGTS beam line and pump ports are part of the IL.
				
				The pumps of the DPS1 reach an ultimate pressure of \SI{<5e-10}{\milli\bar} in the unbaked, \SI{30}{\kelvin} cold WGTS cryostat, without gas load. When gas is circulating, the pumps reduce the gas flow towards the spectrometer by a factor of \num{\approx e3}, as is shown in \autoref{subsec:reduction_factor_inner_loop}.

			\item[Differential Pumping Section 2 (DPS2):] 
				Separated from the DPS1 via a gate valve, four large, cascaded MAG-W2800 TMPs further reduce the downstream flow of neutral tritium in the DPS2 (PP1-4). 
				The fore-vacuum for these pumps is provided by 2 Pfeiffer HiPace300 TMPs. Of these TMPs, the one pumping the PP3 and PP4 MAG-W2800 is cascaded with the other (see \autoref{fig:simplified_loops_schematic}).
				The last HiPace300 TMP is in turn pumped by cascaded fore-pumps, combining a Normetex scroll pump and a metal bellows pump.
				In order to increase the pumping efficiency and prevent a direct line of sight between source and spectrometer, the DPS2 beam tube is arranged in a chicane (see \autoref{fig:simulation_model_dps_beamline}. 
				Gas pumped out from the DPS2 contains a high fraction of outgassing products, which decrease the purity of the tritium gas. Therefore, it is not re-circulated, but returned to the TLK infrastructure for purification. Hence, the DPS2 beam line and pump ports are part of the Outer Loop (OL).
			
				The pumps of the DPS2 reach an ultimate pressure of \SI{<e-9}{\milli\bar} in the unbaked system, without gas load. During gas circulation, the pumps reduce the gas flow towards the spectrometer by a factor of \num{\approx e4}, as is shown in \autoref{subsec:reduction_factor_measurement_outer_loop}.
			
			\item[Cryogenic Pumping Section (CPS):] 
				Separated by another gate valve from the DPS2, about \nicefrac{2}{3} of the CPS beamline is operated at a temperature of \SIrange{3}{4}{\kelvin}, working as a cryo-pump for the remaining tritium. 
				The inner surfaces of the cold beam tubes are enlarged by 90 circular fins welded into the beam tubes and covered by a layer of argon frost. 
				Even with conservative assumptions, simulations indicate a reduction factor of at least \num{e11} \cite{Friedel2019}, well above the minimum design value of \num{e7}. 
				The performance of the CPS is not covered in this paper, but it is used to determine the amount of tritium gas passing the DPS. During the regeneration of the cryo-surfaces the previously sorbed tritium, together with the argon frost, are evaporated and captured in a buffer vessel (B2). 
				The activity of tritium in the gas is measured, allowing the determination of the total amount of tritium gas that passed the DPS.

		\end{description}
		The connections of the IL and OL to the infrastructure systems of the TLK are shown in \autoref{fig:simplified_loops_schematic}.

	\subsection{Definition of reduction factor in existing setup}
		\label{subsec:reduction_factor_definition}

		The overall reduction factor $R_{\mathrm{tot}}$ in the STS denotes the relative reduction of neutral tritium gas flow $Q_{\mathrm{WGTS,d}}$ from the WGTS to the spectrometers. $R_{\mathrm{tot}}$ needs to be larger than \num{e14}. It incorporates the reduction factors of the differential pumping sections $R_{\mathrm{DPS}}$ and of the cryogenic pumping section $R_{\mathrm{CPS}}$, which are both required to reach at least a target value of \num{e7}:
		\begin{equation}
		 R_{\mathrm{tot}} = R_{\mathrm{DPS}} \cdot R_{\mathrm{CPS}}\,.
		\end{equation}

		In addition, $R_{\rm DPS}$ can be subdivided into reduction factors for the inner and outer loops, respectively:
		\begin{equation}
		 R_{\mathrm{DPS}} = R_{\mathrm{IL}} \cdot R_{\mathrm{OL}}\,,
		\end{equation}
		with:
		\begin{description}
		 \item[$R_{\mathrm{IL}}$] is the flow rate reduction between the downstream flow $Q_{\mathrm{WGTS,d}}$ from the injection point into the WGTS to the DPS2. This includes DPS1-F1, DPS1-F2, and PP0 up to V1, as shown in \autoref{fig:simulation_model_dps_beamline}. 
		 \item[$R_{\mathrm{OL}}$] is the flow rate reduction between the flow rate entering PP1 via A3 from the WGTS and the flow rate entering the CPS. In \autoref{fig:simulation_model_dps_beamline} this reduction factor is the fraction of gas entering the DPS2 through V1 over the gas exiting via V2.
		\end{description}
		
		%In previous publications \cite{Kuckert2018,Friedel2019}, $R_{\mathrm{DPS}}$ was subdivided for simulation purposes. The partial reduction factors which are used in the context of this work are slightly different from the reduction factors defined by previously done simulations. This is due to the fact that the simulations have split the different reduction factors by different flow regimes and accordingly different modelling.
		%For the experimental determination of the reduction factors however, the measurable quantities are segmented by whether they belong to the IL or the OL:
		
		In previous simulations \cite{Kuckert2018,Friedel2019} $R_{\mathrm{DPS}}$ was subdivided according to the different flow regimes, using laminar and transitional flow up to DPS1-2F and molecular flow in PP0 and DPS2. However, for the measurements the reduction factors have to be split into $R_{\mathrm{IL}}$ and $R_{\mathrm{OL}}$.

% 		\color{red}{Use \autoref{fig:simulation_model_dps_beamline} in this section}
% 		\color{blue}
% 		\begin{itemize}
% 			\item Give definitions for the reduction factors used in this paper
% 			\begin{itemize}
% 				\item $R_{\mathrm{Inner Loop}}$ is the reduction in detector bound flowrate from the injection point into the WGTS to the flow rate entering the DPS2. This encompasses DPS1-F1, DPS1-F2, and PP0 up to \textbf{V1} as shown in \autoref{fig:simulation_model_dps_beamline}. 
% 				\item $R_{\mathrm{Outer Loop}}$ is the reduction in flow rate between the incoming flow rate into PP1 from the source and the flow rate entering the CPS. In \autoref{fig:simulation_model_dps_beamline} this reduction factor is the fraction of gas entering the DPS2 through \textbf{V1} over the gas existing via \textbf{V2}.
% 			\end{itemize}
% 			\item Explicitly explain difference in terminology to simulation paper
% 			\begin{itemize}
% 				\item In \cite{Kuckert2018} a value for only DPS1-F1 and DPS1-F2 was giving, without PP0
% 				\item In \cite{Friedel2019} a value for the DPS consisting of PP0-PP5 was given.
% 			\end{itemize}
% 			\item Explain difference to simulated reduction factors. Simulations contain PP0 different from what can be experimentally determined.
% 			
% 			\item Try to keep definitions as they are done in Florian Priester's Paper for the Tritium 2019 conference
% 		\end{itemize}
% 		\color{black}

	\subsection{Results of simulations}
		\label{subsec:results_of_simulations}
	
		\begin{figure}
		\centering
			\includegraphics[width=\textwidth]{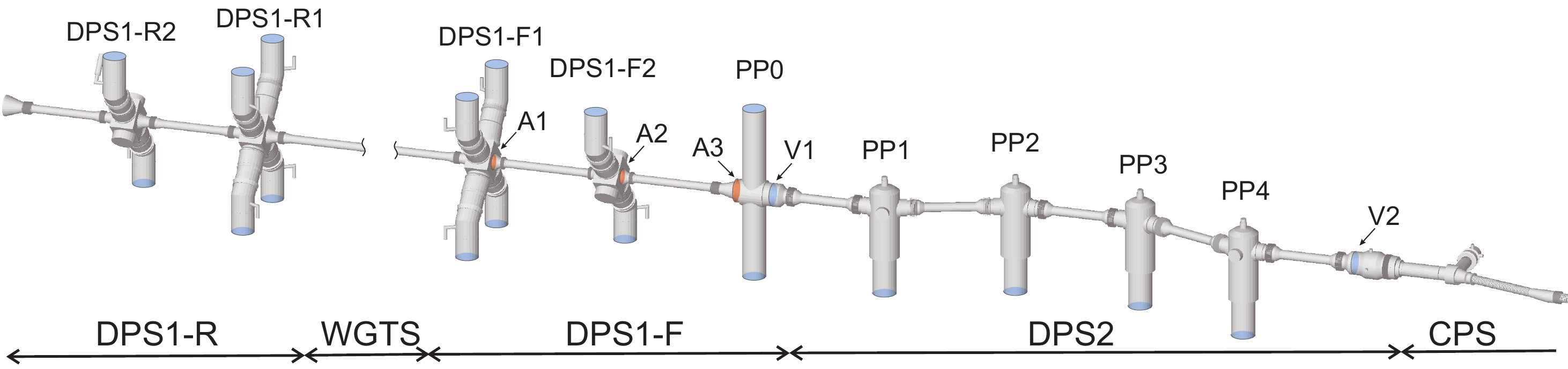}
			\caption[KATRIN beamline overview]{{Schematic drawing of the differential pumping section of the beam line.} 
			It shows only the beam tubes and the pump ports of the vacuum system, omitting the enclosing superconducting magnets and cryostats. The valve V1 separates the differential pumping sections connected to either the Inner Loop or the Outer Loop.}
			\label{fig:simulation_model_dps_beamline}
		\end{figure}	
	
	%Two sets of simulations were done for the DPS in prior works \cite{Kuckert2018,Friedel2019}.
	
	The DPS1 is described in Ref. \cite{Kuckert2018}, which splits it into three computational domains. For the transition from the laminar flow regime at the injection point towards the transitional flow regime in the first pump ports, a semi-analytical rarefied gas dynamics model \cite{Sharipov2016,Sharipov1997} is used. The transitional flow regime inside DPS1-F1 is simulated using a Direct Simulation Monte Carlo (DSMC) \cite{Bird1994} method. Finally, the outermost pump ports are described, using the angular coefficient \cite{Kersevan2009} method to account for the transition from \SI{30}{\kelvin} to room temperature in this domain.
	
	As the focus in \cite{Kuckert2018} was on the precise description of the gas density distribution along the beamline, the reduction factor was not investigated in detail, and therefore no detailed error analysis was given. From the uncertainties given on the simulated pressures and flows, we estimate an uncertainty of about a factor two in both directions, mainly owing to the uncertainty of the effective pumping speed used in the simplified geometry of the model. The value for the gas flow reduction as reported in Ref. \cite{Kuckert2018} is:
	\begin{equation}
	\label{eq:kuckert-result}
	 R_{\mathrm{DPS1}}^{\mathrm{Sim}} = \SI[parse-numbers=false]{386^{+386}_{-193}}\,.
	\end{equation}
	In order to achieve comparability between measurement and simulation, the DPS2 MolFlow+ simulation performed in \cite{Friedel2019} has been rerun with slightly different boundary conditions, since the intitial simulation included PP0 as part of the DPS2, while in the measurements it is connected to the IL.
	
	MolFlow+ \cite{Kersevan2009} is a Test Particle Monte Carlo (TPMC) simulation code for particle tracking through the geometry of a vacuum chamber in the molecular flow regime.
	Particles do not interact with each other but only with the walls of the vacuum chamber where they get adsorbed, desorbed or reflected.
	The geometry is approximated by a mesh of two-dimensional polygon surfaces, called facets.
	For each facet the number of hits, adsorptions and desorptions is counted. 	
	An adsorbing facet represents a pump, a desorbing facet a gas source.	
	Fully transparent (virtual) facets which are not part of the actual physical geometry can be defined providing additional counting of hits at a location of interest.
	Ratios of counts are used in order to determine transmission probabilities.
	In \autoref{fig:simulation_model_dps_beamline} an overview of the implemented geometry is given. Besides the physical boundaries of the beam line there are several virtual facets implemented. The particle tracking starts at facet A2 downstream of DPS1-F2. Particles are removed from the simulation in three different cases:
	They
	\begin{itemize}
		\item hit facet A1, 
		\item are pumped out by one of the TMPs in the pumping ports DPS1-F2 or PP0-4, or
		\item are pumped out at the CPS cryo-pump downstream of V2.
	\end{itemize} 
	Consequently, the simulation does not only consist of the DPS2, but also the neighboring sections. 
	Since particle tracking starts already at A2, while only facet counts at A3 and beyond are used in the simulation, it is assured, that boundary effects, such as back reflections or the angular distributions of particle velocities are included correctly. In MolFlow+ pumps are modeled by facets with well-defined sticking probabilities $\alpha \in [0,1]$, corresponding to the pump's gas type dependent pumping probability. For the CPS cryo-pump $\alpha=0.7$ was set, which is an established reference value of a well prepared argon layer at 3\,K \cite{Yuferov1970}. Particles moving as far back as DPS1-F1, hitting facet A1, are assumed to be pumped off. Consequently, these particles are removed from the simulation by setting $\alpha_{\mathrm{A1}}=1$. It has been verified that this simplification of the model does affect the results of the simulation by less than \SI{0.5}{\percent}. 
	
	The DPS TMPs were included with $\alpha=\num{0.252}$, corresponding to their estimated pumping probability for particles of mass $m=\SI{6}{\gram\per\mole}$. For this estimate we used the nominal pumping speeds given by the manufaturer, interpolating different particle masses by applying the Malyshev model~\cite{Malyshev2007}, which assumes that the pumping probability scales with the logarithm of the particle mass $M$ ($\alpha \propto \ln(M)$). The systematic uncertainty of this method has been taken into account as an uncertainty of \SI{20}{\percent} on this pumping probability. It was estimated by comparing the measured and simulated pumping speeds for different gases (based on the nominal pumping speed of the pump manufacturer) in the KATRIN Main Spectrometer. The impact on the resulting reduction factor was obtained by dedicated simulations with \SI{20}{\percent} higher and \SI{20}{\percent} lower TMP pumping probabilities, respectively.

In table \ref{tab:molflowresults} the output of the simulations is shown by giving the important numbers for the reduction factor calculations. 
Using the number of particles pumped at PP1-4 ($N_{\mathrm{PP1-4}}$) and at the CPS ($N_{\mathrm{CPS}}$) for $\alpha=\num{0.25}$ the resulting reduction factor is derived by the following equation:
\begin{equation}
\label{eq:R_sim_ol}
R_{\mathrm{OL}}^{\mathrm{Sim}} = \frac{N_{\mathrm{PP1-4}}}{N_{\mathrm{CPS}}}=
	\num[parse-numbers=false]{1.50^{+0.69}_{-0.58}\times 10^{4}} .
\end{equation}
The upper and lower uncertainties originate from the simulations with $\alpha=\num{0.20}$ and $\alpha=\num{0.30}$ respectively. 
Since the maximal statistical uncertainty is about \SI{4}{\percent} and thus much smaller than the systematic uncertainty of about \SI{\approx 40}{\percent}, it is neglected in the following.
\begin{table}
	\centering
	\caption[Results of the DPS2 gas flow simulations with MolFlow+.]{\textbf{Results of the DPS2 gas flow simulations with MolFlow+.} Shown are the number of simulated test particles. Additionally, the numbers of particles terminated either at a group of TMPs or at the CPS cryo-pump are listed. These values are given for different TMP pumping probabilities $\alpha$. For the number of particles entering PP0 via A3 ($N_{\rm A3}$), back reflection through A3 towards DPS1-F2 has been taken into account by subtracting the hits of particles through A3 in upstream direction from the number of particles entering in downstream direction. This is equivallent to the total number of particles pumped by all adsorbing facets downstream of A3.}
	\label{tab:molflowresults}
	%\resizebox{\linewidth}{!}{
	%\begin{tabular}{cccc}
	\begin{tabularx}{\textwidth}{p{0.38\textwidth}XXXX}
		\toprule
		Particle tagged as & Notation & $\alpha = \num{0.20}$ & $\alpha = \num{0.25}$ & $\alpha = \num{0.30}$  \\ \midrule 
		Total simulated particles & & \num{2.07e9} & \num{2.13e9} & \num{2.21e9} \\ 
		Entering PP0 via A3 & $N_{\mathrm{A3}}$ & \num{1.55e8} & \num{1.49e8} & \num{1.48e8} \\
		Pumped at PP0 & $N_{\mathrm{PP0}}$ & \num{1.43e8} & \num{1.39e8} & \num{1.39e8} \\ 
		Pumped at PP1-4 & $N_{\mathrm{PP1-4}}$ & \num{1.16e7} & \num{1.02e7} & \num{9.42e6} \\ 
		Pumped at CPS &	$N_{\mathrm{CPS}}$ & \num{1.26e3}& \num{6.81e2} & \num{4.31e2} \\ 
		\bottomrule
	\end{tabularx}
	%\end{tabular}
%	}

\end{table}
 
%\textcolor{red}{[Still missing: Combine Kuckert and Friedel results here, give $R_{\mathrm{Overall}}^{\mathrm{Simulation}}$]}

The reduction factor of the IL can be derived from $R_{\mathrm{DPS1}}^{\mathrm{Sim}}$ in \autoref{eq:kuckert-result} and the reduction factor of PP0.  
With the simulated numbers from table 1, the reduction of the gas flow via PP0 is given by the ratio of particles entering PP0 through A3 ($N_{\mathrm{A3}}$) and those pumped by the DPS2 ($N_{\mathrm{PP1-4}}$) and the CPS ($N_{\mathrm{CPS}}$):

\begin{equation}
 R_{\mathrm{PP0}}^{\mathrm{Sim}} = \frac{N_{\mathrm{A3}}}{N_{\mathrm{PP1-4}}+N_{\mathrm{CPS}}} = \num[parse-numbers=false]{1.460^{+0.110}_{-0.120}\times 10^{1}}\,.
\end{equation}

\noindent As for $R_{\mathrm{DPS1}}^{\mathrm{Sim}}$ however, there is one difference between the setup for which $R_{\mathrm{DPS1}}^{\mathrm{Sim}}$ was simulated and how it is operated in the current KATRIN setup.
In \cite{Kuckert2018} the simulation ended at the surface A3 with an assumed effective pumping probability of $\alpha = \SI{20}{\percent}$ for PP0 and the subsequent DPS2 pumps.
This number originated from calculations for the case of only one active TMP at PP0. 
Currently, both TMPs are operated.
So, the A3 effective pumping probability, as simulated with MolFlow+, should be rather \SI{36}{\percent} than \SI{20}{\percent}. 
The impact on the final result of $R_{\mathrm{DPS1}}^{\mathrm{Sim}}$ has been calculated based on two dedicated MolFlow+ simulations.
In each of them A3 is assumed as an opaque facet but with different sticking factors of \SI{20}{\percent} and \SI{36}{\percent}, respectively.
Gas particles are desorbed from facet A1. 
For both simulations the reduction factors have been calculated by taking the ratio of the number of adsorptions at A3 and hits at A2. 
The correction factor $C_{\mathrm{A3}}$ for $R_{\mathrm{DPS1}}^{\mathrm{Sim}}$ is determined as the ratio of both reduction factors, resulting in a value of ${C_{\mathrm{A3}} = R_{\SI{36}{\percent}}/R_{\SI{20}{\percent}} = \num{0.953 \pm 0.003}}$. 
The corrected IL reduction factor is:

\begin{equation}
 R_{\mathrm{IL}}^{\mathrm{Sim}} = C_{\mathrm{A3}} \cdot R_{\mathrm{DPS1}}^{\mathrm{Sim}} \cdot R_{\mathrm{PP0}}^{\mathrm{Sim}} = \num[parse-numbers=false]{5.4^{+6.2}_{-2.9}\times 10^{3}}\,.
\end{equation}

\noindent In combination with the simulated OL result from equation \ref{eq:R_sim_ol} one can derive the overall simulated reduction factor for the DPS:

\begin{equation}
 R_{\mathrm{DPS}}^{\mathrm{Sim}} = R_{\mathrm{IL}}^{\mathrm{Sim}} \cdot R_{\mathrm{OL}}^{\mathrm{Sim}} = \num[parse-numbers=false]{8^{+17}_{-6}\times 10^{7}}\,.
\end{equation}

\noindent With the assumption that the uncertainties of the TMP pumping probabilities in DPS1 and DPS2 are correlated, the uncertainties were estimated by multiplying the simulations with the upper and lower bounds of the two reduction factors, respectively, and subtracting the simulation with the central value. 

% updated: 10.07.2020, J.W.

\section{Measurement of reduction factors}
	\label{sec:measurement}
% 	Use \autoref{fig:simplified_loops_schematic} in this section to explain the measurement methods. Probably remake the graphic and remove unnecessary parts.

	\sloppy
	Two different methods were used to determine the reduction factors $R_{\mathrm{IL}}$ and $R_{\mathrm{OL}}$ for tritium. Measurements were taken for the nominal column density of ${\SI{5.0e21}{\per\square\meter}\, (\estimates \SI{100}{\percent}})$ as well as for the settings used during the KNM1 $({\SI{1.1e21}{\per\square\meter}\, \estimates \SI{22}{\percent}})$ and KNM2 $({\SI{4.2e21}{\per\square\meter}\, \estimates \SI{84}{\percent}})$ measurement campaigns. 
	While the reduction factor $R_{\mathrm{IL}}$ depends on the pressure in the WGTS beam tube and the temperature, $R_{\mathrm{OL}}$ is constant, since the DPS2 is operated at constant room temperature in the molecular flow regime.
	In contrast to the upper and lower bounds of uncertainty present for the theoretical results, the uncertainties for the derived quantities were calculated using uncertainty propagation assuming gaussian distributed uncertainties of the measured values.
	The different methods and their results are described below.
	
% 	
% 	\begin{figure}
% 	\centering
% 		\includegraphics[width=\textwidth]{../figures/simplified_flow_diagram.png}
% 		\caption[Simplified Flow Diagram]{\textbf{Simplified flow diagram of the beamline and its pumps.} The TMPs at WGTS and DPS, their piping, as well as the CPS waste buffer should be shown here to be able to explain the measurements. This image should contain only very schematic depictions for the sake of clarity when explaining the measurements.}
% 		\label{fig:simplified_loops_schematic}
% 	\end{figure}

\subsection{Reduction factor of the Inner Loop}
	\label{subsec:reduction_factor_inner_loop}

	The reduction factor $R_{\mathrm{IL}}$ is determined from the ratio of the measured gas flow rates into the WGTS and into the DPS2:
	
	\begin{equation}
			R_{\mathrm{IL}} = \frac{Q_{\mathrm{WGTS,d}}}{Q_{\mathrm{DPS2}}} \,.
	\end{equation}
			
	\noindent The gas flow rate in downstream direction $Q_{\mathrm{WGTS,d}}$ inside the WGTS is calculated using a {MKS 179} mass flow meter\footnote{Full Scale (F.S.) of \SI{200}{sccm} (\SI{\approx 3.4}{\milli\bar\litre\per\second}), accuracy \SI{1}{\percent} of F.S., \SI{\approx 2}{\percent} at nominal column density.} (labeled FIR in \autoref{fig:simplified_loops_schematic}), which measures the total flow rate into the WGTS $Q_{\mathrm{WGTS,tot}}$. With the symmetric design of the WGTS beam tube and the DPS1-R and DPS1-F pump ports, equal conductances and effective pumping speeds in upstream and downstream direction can be assumed, leading to an equal split of the flow rate in both directions:

		\begin{equation}
			Q_{\mathrm{WGTS,d}} = \frac{1}{2} \cdot Q_{\mathrm{WGTS,tot}} \,.
			\label{eq:wgts_flow}
		\end{equation}

	\noindent The pressure ratio between the point of injection and both ends of the symmetrical sections, DPS1-F2 and DPS1-R2, is around \num{3} orders of magnitude. Therefore, the small effect of the difference in effective pumping speeds between the Rear Section at the upstream end and DPS2-PP0 at the downstream end can be neglected, compared to the systematic uncertainties of flow and pressure measurements which are on the percent-level. 

	The effective gas flow rate $Q_{\mathrm{DPS2}}$ entering the DPS2 is measured by a pressure rise $\nicefrac{\Delta p}{\Delta t}$ inside the buffer vessel B1 (labeled PIR in \autoref{fig:simplified_loops_schematic}) with a well known volume $V_{\mathrm{B1}}$, located behind the last cascaded DPS2 TMP:

		\begin{equation}
			Q_{\mathrm{DPS2}} = V_{\mathrm{B1}} \cdot \frac{\Delta p}{\Delta t} \,.
			\label{eq:dps_flow}
		\end{equation}

	\noindent The volume $V_{\mathrm{B1}} = \SI{16.53\pm0.21}{\litre}$ has been determined during the commissioning phase via gas expansion from a reference volume. 
	A necessary assumption for applying \autoref{eq:dps_flow} is that all gas entering the DPS2 is pumped out by the \num{4} TMPs connected to the beamline, neglecting the gas entering the CPS.
	This assumption is justified, since the flow rate into the CPS is reduced by four orders of magnitude in the DPS2, which is much smaller compared to the systematic uncertainties for the flow and pressure measurement in the percent-level (see  \autoref{subsec:reduction_factor_measurement_outer_loop}).

	An additional effect is the outgassing of the DPS2 setup. The surfaces of the vacuum chambers, beam line instrumentation such as dipole electrodes or an ion monitor inside the vacuum system of the DPS2, as well as the TMPs themselves cause a non-negligible outgassing rate, leading to an additional pressure rise $\nicefrac{\Delta p_{\mathrm{og}}}{\Delta t_{\mathrm{og}}}$ in $V_{\mathrm{B1}}$. 
	In order to correct for this effect, the outgassing rate was measured during operation of the beam line without tritium gas injection and then subtracted from the tritium gas flow induced pressure rise:

		\begin{equation}
			Q_{\mathrm{DPS2}} = V_{\mathrm{B1}} \cdot \left( \frac{\Delta p}{\Delta t} - \frac{\Delta p_{\mathrm{og}}}{\Delta t_{\mathrm{og}}}\right) \,.
		\end{equation}
		\label{subsubsec:wgts_results}
			
\begin{figure}
		\centering
			\includegraphics[width=\linewidth]{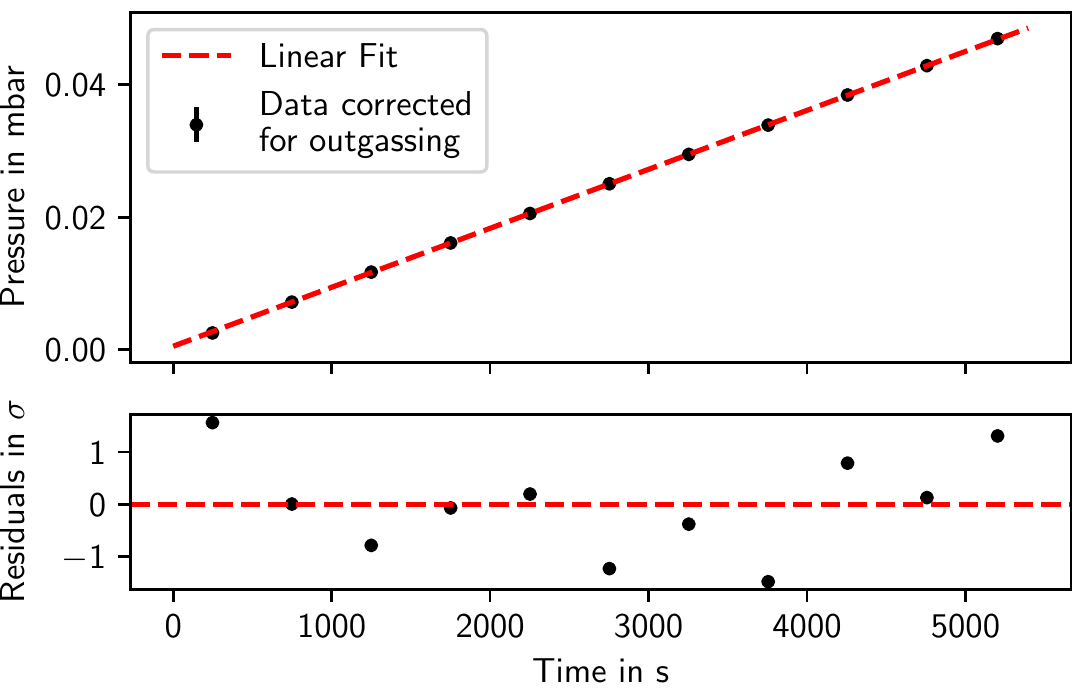}
			\caption[Pressure rise fit for the gas flow rate comparison measurement]{
			{Pressure in the buffer vessel B1 connected to the DPS2 TMPs vs. time.} 
			Shown as an example is the measurement with a column density of ${\SI{5.0e21}{\per\square\meter} (\estimates \SI{100}{\percent}})$. Measurements at other column densities show the same behavior with the only difference being the slope of the linear function. The outgassing rate was for all measurements about \SI{5e-7}{\milli\bar\litre\per\second}, which is only \SI{1}{\percent} of $Q_{\mathrm{DPS2}}$ at \SI{100}{\percent} column density.}
			\label{fig:pressure_rise_dps_loops_fit}
		\end{figure}

	\noindent The outgassing rate ($\approx$\,\SI{5e-7}{\milli\bar\litre\per\second}) is determined for each tritium measurement from the latest available outgassing measurement, in order to account for possible changes in the outgassing behavior. 
	The pressure rise data and a linear fit for a column density of \SI{5.0e21}{\per\square\meter} can be seen in \autoref{fig:pressure_rise_dps_loops_fit}. 
	The results for $Q_{\mathrm{DPS2}}$, $Q_{\mathrm{WGTS,d}}$, and $R_{\mathrm{IL}}$ for different column densities are listed in \autoref{table:pressure_rise_results}.

		\begin{table}
			\caption[Results for the gas flow rate comparison measurement]{\textbf{Results for the Inner Loop reduction factor measurements at different column densities.} 
			Given are the values for the gas flow rate from the center of the WGTS in downstream direction $Q_{\mathrm{WGTS,d}}$, fit results for the effective gas flow rate $Q_{\mathrm{DPS2}}$ entering the DPS2, as well as the Inner Loop reduction factors ${R_{\mathrm{IL}} = {Q_{\mathrm{WGTS,d}}}/{Q_{\mathrm{DPS2}}}}$. 
			The identical uncertainties for $Q_{\mathrm{WGTS,d}}$ are due to the constant \SI{1}{\percent} F.S. accuracy of the flow meter.
			The uncertainty of $Q_{\mathrm{DPS2}}$ is dominated by the uncertainty of the buffer volume $\delta V_{\rm B1}/V_{\rm B1} = \SI{1.3}{\percent}$. The uncertainties of the pressure measurements at the buffer vessel are \SI{0.5}{\percent}. For $R_{\mathrm{IL}}$ the errors are propagated in quadrature.}
			\label{table:pressure_rise_results}
			\centering
			\begin{tabularx}{\textwidth}{XXXXX} \toprule
				Column Density  & Temperature & $Q_{\mathrm{WGTS,d}}$  & $Q_{\mathrm{DPS2}}$  & $R_{\mathrm{IL}}$ \\ 
					\SI{e21}{\per\square\meter} & \si{\kelvin} & \si{\milli\bar\litre\per\second} & \SI{e-4}{\milli\bar\litre\per\second} & \num{e3}\\ \midrule
				\num{1.1}  & \num{30}  & \num{0.112\pm0.017} & \num{0.293\pm0.004} & \num{3.82\pm0.58} \\
				\num{1.8}  & \num{80}  & \num{0.256\pm0.017} & \num{0.627\pm0.009} & \num{4.09\pm0.28} \\
				\num{2.0}  & \num{100} & \num{0.354\pm0.017} & \num{0.782\pm0.010} & \num{4.53\pm0.22} \\
				\num{4.2}  & \num{30}  & \num{0.756\pm0.017} & \num{1.159\pm0.015} & \num{6.52\pm0.17} \\
				\num{5.0}  & \num{30}  & \num{0.982\pm0.017} & \num{1.471\pm0.019} & \num{6.68\pm0.14} \\ \bottomrule
			\end{tabularx}
		\end{table}
		
	The monitoring of $R_{\mathrm{IL}}$ with resonable accuracy has two direct applications for the operation of the source and transport section. 
	First, with a constant $R_{\mathrm{OL}}$ the expected tritium load on the CPS is accessible with a measurement on the time scale of \SIrange{1}{2}{\hour}. This allows for reduction factor measurements to be done for different settings of WGTS cryostat temperatures and gas flows, which can change $R_{\mathrm{IL}}$. 
	Second, the fraction of gas which can be recirculated in the IL can be directly derived from this measurement. 
	As such, $R_{\mathrm{IL}}$ has a direct impact on the operation of both the IL and OL.

\subsection{Reduction factor of the Outer Loop}
	\label{subsec:reduction_factor_measurement_outer_loop}
		\sloppy
		%The magnitude of the $R_{\mathrm{OL}}$ and the associated flow $Q_{\mathrm{CPS}}$ into the CPS has direct implications for the time between two regenerations of the CPS argon frost layer. Thus, by determining $R_{\mathrm{OL}}$, the time span of a neutrino mass measurement run can be assessed.
		%The capacity limit of the CPS is in this case not given by the gas dynamical limitations and saturation of the cryogenic surface, but by the accumulated tritium activity bound onto the pumping surface. Due to radiation safety concerns, this is limited to ${A_{\mathrm{CPS,max}} = \SI{3.7e10}{\becquerel}}$. This activity corresponds to ${N_{\mathrm{CPS,max}} = \SI{0.39}{\milli\bar\litre}}$ of T$_2$ gas at standard conditions.

		The OL reduction factor $R_{\mathrm{OL}}$ is calculated in a two step process. First, a combined reduction factor $R_{\mathrm{DPS}} = R_{\mathrm{IL}} \cdot R_{\mathrm{OL}}$ is measured by comparing integral gas activities:
		\begin{equation}
			R_{\mathrm{DPS}} = \frac{A_{\mathrm{WGTS}}}{A_{\mathrm{CPS}}} \,.
		\end{equation}
		Where $A_{\mathrm{WGTS}}$ is the integral activity of beta-decays in the gas flow $Q_{\mathrm{WGTS,d}}$, and $A_{\mathrm{CPS}}$ is the beta-activity accumulated inside the CPS. The direct relation of the accumulated beta-activity to the integral gas flow into the CPS can be made, as isotopic exchange effects inside the DPS2 are expected to be on the sub-percent-level and therefore insignificant. This expectation is derived from the IL gas composition measurements for which gas passes through WGTS and DPS1, which are comparable to the DPS2 in length, the composition changes are below the percent-level for a single pass through.
		
		To obtain $R_{\mathrm{OL}}$, $R_{\mathrm{DPS}}$ is divided by $R_{\mathrm{IL}}$:
		\begin{equation}
			R_{\mathrm{OL}} = \frac{R_{\mathrm{DPS}}}{R_{\mathrm{IL}}} = \frac{1}{R_{\mathrm{IL}}} \cdot \frac{A_{\mathrm{WGTS}}}{A_{\mathrm{CPS}}} \,.
			\label{eq:outer_loop_reduction_factor}
		\end{equation}
		
		\noindent This measurement method, using the activity of tritium gas, is needed as the cryogenic pumping principle of the CPS does not allow for an easily measurable gas accumulation in situ as described in \autoref{subsec:reduction_factor_inner_loop}. 
		The gas flow entering the CPS is adsorbed on its cryogenic surface and can only be determined after regeneration. 
		During the regeneration procedure, helium is used as a purge gas to remove the argon frost layer together with the captured tritium.
		This results in a mixture of around \SI{\sim 6}{\bar\litre} argon, \SI{250}{\bar\litre} helium, and traces of tritium of less than \SI{0.39}{\milli\bar\litre}, limited by the maximum allowed activity inside the CPS. 
		The entire gas is collected in the buffer vessel B2 (see \autoref{fig:simplified_loops_schematic}).
		Reliable quantification of the small trace amounts of tritium in this gas mixture is impossible using pressure measurements and very challenging using residual gas analyzers. 
		However, the traces of tritium are quantifiable by counting the beta-activity in the gas with measurement techniques developed by TLK \cite{Corneli2005,Roellig2015,Ebenhoech2016}. 
		
		%After regeneration of the CPS, any gas containing tritium is purged and then collected in a buffer vessel (B2 in \autoref{fig:simplified_loops_schematic}).
		Several samples of this gas mixture from the buffer vessel were analyzed, using oxidation on copper oxide (CuO) at \SI{450}{\celsius}, followed by liquid scintillation counting to determine the activity concentration of the gas sample with an uncertainty of \SI{10}{\percent}. 
		%\textcolor{red}{{\bf Marco:} check comment L335 and update this paragraph; which method was used? should we mention BIXS or not?, What is the uncertainty?}
		%Additionally, in-situ measurements with a beta-induced X-ray spectroscopy (BIXS) setup are conducted at the buffer vessel. 
       % \textcolor{red}{Could we write: For future in-situ monitoring of the accumulated activity in B2, the BIXS results have been cross-calibrated... ?}
		%They are than cross-calibrated with the results of the oxidation measurements. 
		%For the results reported here the oxidation method was used to determine the total activity of the collected purge gas $A_{\mathrm{CPS}}$.
		The total activity of the collected purge gas $A_{\mathrm{CPS}}$ was calculated by scaling the sample activity with the respective gas amounts.

		The determination of the activity $A_{\mathrm{WGTS}}$ is not possible via a direct activity measurement. 
		It can be derived from the gas flow $Q_{\mathrm{WGTS,d}}$ and the composition of the gas. 
		The gas composition is measured via laser raman spectroscopy \cite{Sturm2009, Schloesser2013, Aker2020a}. 
		The composition analysis allows the determination of the fraction of tritium $\epsilon_\mathrm{T}$. 
		Using these two values and the specific activity $a_{\mathrm{T}_2} = \SI{9.5e10}{\becquerel/\milli\bar\litre}$ of T$_2$, one can calculate the integral activity $A_{\mathrm{WGTS}}$ as follows:

		\begin{equation}
			A_{\mathrm{WGTS}} = a_{\mathrm{T}_2} \cdot \int{Q_{\mathrm{WGTS,d}}(t) \cdot \epsilon_\mathrm{T}(t) \dif t} \,.
		\end{equation}
		%For the purpose of this reduction factor measurements, only those times were taken into account, in which gas flow into the CPS was possible.
		
		\noindent The measurement results for $A_{\mathrm{WGTS}}$ and $A_{\mathrm{CPS}}$ for the CPS regenerations after the KNM1 and KNM2 measurement campaigns, as well as the reduction factors $R_{\mathrm{DPS}}$ and $R_{\mathrm{OL}}$, are shown in \autoref{table:integral_activity_results}.
	
		\begin{table}
			\caption[Results for the integral activity measurement]{Results for the combined reduction factor $R_{\mathrm{DPS}}$ measurements during the KNM1 and KNM2 measurement campaigns. 
			Given are the values for integral beta-activity $A_{\mathrm{WGTS}}$ of the WGTS downstream gas flow $Q_{\mathrm{WGTS,d}}$, the total accumulated beta-activity on the CPS $A_{\mathrm{CPS}}$, the combined reduction factor $R_{\mathrm{DPS}}$ as well as the Outer Loop reduction factor $R_{\mathrm{OL}} = A_{\mathrm{WGTS}} \cdot A_{\mathrm{CPS}}^{-1} \cdot R_{\mathrm{IL}}^{-1}$ derived using the Inner Loop reduction factors $R_{\mathrm{IL}}$ from \autoref{table:pressure_rise_results}.}
			\label{table:integral_activity_results}
			\centering
			\begin{tabularx}{\textwidth}{XXXXX} \toprule
				Measurement & $A_{\mathrm{WGTS}}$ & $A_{\mathrm{CPS}}$ & $R_{\mathrm{DPS}}$ & $R_{\mathrm{OL}}$ \\ 
				Campaign & in \SI{e17}{\becquerel} & in \SI{e9}{\becquerel} & in \num{e7} & in \num{e4} \\ \midrule
				KNM1  & \num{0.94\pm0.13} & \num{1.56\pm0.16} & \num{6.04\pm1.03} & \num{1.58\pm0.36}\\
				KNM2  & \num{4.10\pm0.12} & \num{4.26\pm0.43} & \num{9.63\pm1.00} & \num{1.48\pm0.16}\\ \bottomrule
			\end{tabularx}
		\end{table}
	
% 		
% 		\begin{figure}
% 		\centering
% 			\includegraphics[width=\linewidth]{../figures/cumulative_throughput.png}
% 			\caption[Throughput data for the KNM1 measurement phase.]{\textbf{Throughput measurement data used for calculating the integrated activity comparison measurement.} Shown is the measured injection flow rate into the WGTS as well as its cumulative. The cumulative value, in combination with the tritium purity and specific activity, gives an integral activity which can be compared with the activity measured in the exhaust gas gained from CPS regeneration.}
% 			\label{fig:cumulative_throughput_measurement}
% 		\end{figure}

\section{Discussion}
\label{sec:discussion}
% 	\color{blue}
% \begin{itemize}
%  \item Discuss difference between simulation and measurement results.
%  \item Important to look at the error bars to correctly interpret any differences.
%  \item Evaluate the meaning of the measured reduction factor for KATRIN
%  \begin{itemize}
%   \item Impact on CPS runtime
%   \item Connect to CPS reduction measurements with deuterium (Carsten)
%  \end{itemize}
%  
% \end{itemize}
% 	\color{black}	

	%The agreement between the measured reduction factors for the IL and the OL, with the previously simulated values shown in \autoref{subsec:results_of_simulations}, will be discussed below. 
	%In the following, a comparison between the measured and the simulated reduction factors is made for $R_{\mathrm{IL}}$ and $R_{\mathrm{OL}}$ respectively. 
	%Additionally, the runtime limitations of the cryogenic pumping section will be discussed in the context of the measured reduction factor. 

	\subsection{Inner Loop reduction factor}	
		 
		The simulated IL reduction factor for the nominal column density of ${\SI{5e21}{\per\square\meter} ({\estimates \SI{100}{\percent}})}$ as derived in this work (see \autoref{subsec:results_of_simulations}) is:

		\begin{equation}
		%R_{\mathrm{IL}}^{\mathrm{Sim}} = \num[parse-numbers=false]{5.6^{+0.5}_{-0.4}\times 10^{3}} .
		R_{\mathrm{IL}}^{\mathrm{Sim}} = \num[parse-numbers=false]{5.4^{+6.2}_{-2.9}\times 10^{3}} .
		\end{equation}

		\noindent Comparing this to the measured value of the IL reduction factor,
		
		\begin{equation}
		R_{\mathrm{IL}} = \num{6.68\pm0.14e3} ,
		\end{equation}
		
		\sloppy 
		\noindent one can see that the measured value is close to the simulated value for the central value of the pumping probability $\alpha=\num{0.25}$, and well within the uncertainty of the simulation. 
		%one can see that the measured value is \SI{\approx 16}{\percent} higher than the simulated value, and their respective uncertainties do not overlap. 
		%However, if one assumes for the missing uncertainty on the reduction factor from \cite{Kuckert2018} reasonable values of \SIrange{10}{20}{\percent}, this discrepancy vanishes.
		
		An effect which had not been expected initially, is the significantly different reduction factor at the low column density setting of ${\SI{1.1e21}{\per\square\meter} ({\estimatesapprox \SI{22}{\percent}})}$ used during KNM1 (see \autoref{table:pressure_rise_results}). 
		This strong dependence of the reduction factor on the column density, and thereby pressure and flow, can be attributed to the changes of flow regime inside the WGTS. 
		With decreasing column density, the pressure inside the pump ports decreases, shifting the flow regime from the Knudsen flow regime further towards the free molecular flow regime. 
		In the Knudsen flow regime more scattering of gas molecules inside the DPS1-F1 is present. 
		The narrow and long geometry of the beam line between the injection point and DPS1-F1 produces a distinct molecular beam. 
		Thus, radial movement is suppressed and molecules only receive a strong radial momentum by scattering with other molecules. 
		Since gas particles are only pumped if they move radially towards the TMPs, less scattering produces lower reduction factors. 
		This effect is the most likely reason for the smaller reduction factors of the IL at low column densities.
		Various measurements of $R_{\mathrm{IL}}$ at different column densities and beam line temperatures showed that there is no strong influence of the temperature, but a clear correlation with the column density (see \autoref{table:pressure_rise_results}).
		As the DSMC simulations for the Knudsen flow regime are computationally intensive, and low column densities are not of interest for normal KATRIN operation, a detailed parameter study was not undertaken.

	\subsection{Outer Loop reduction factor}	
		The simulated OL reduction factor, as derived in this work (see \autoref{subsec:results_of_simulations}), is:
		\begin{equation}
		R_{\mathrm{OL}}^{\mathrm{Sim}} = \num[parse-numbers=false]{1.50^{+0.69}_{-0.58}\times 10^{4}} .
		\end{equation}
		
		Comparing this to the value of the OL reduction factor derived in the KNM2 measurement campaign,

		\begin{equation}
		R_{\mathrm{OL}} = \num{1.48\pm0.16e4} ,
		\end{equation}

		\noindent one can see that the values match well within their respective uncertainties. 
		In contrast to $R_{\mathrm{IL}}$, no dependance of $R_{\mathrm{OL}}$ on the column density can be inferred from the data, considering the measurement uncertainties.
		This is in good agreement with the underlying assumption of free molecular flow inside the OL section of the differential pumping section. 
	
		%Also, these findings are in reasonable agreement with stand-alone measurements on the predecessor of the current DPS2-F, where the geometry was slightly different. 
		%A retention factor of \SI{2.5e4} for Tritium was predicted. /LUC2012/. At that time only non-radioactive gases could be used, so that the tritium value had to be assessed by interpolation.

	\subsection{Impact of the reduction factor on CPS operation}	
		\sloppy 
		While there is no data on the combined reduction factor for the differential pumping sections $R_{\mathrm{DPS}}$ at nominal column densitiy of \SI{5e21}{\per\square\meter}, an estimation can be made using the data gained from KNM2 with a column density of \SI{4.2e21}{\per\square\meter}. The difference between $R_{\mathrm{IL}}$ for both column densities is negligable, and $R_{\mathrm{OL}}$ does not depend on the column density value.
		As such, the $R_{\mathrm{DPS}}$ measured during KNM2,

		\begin{equation}
		R_{\mathrm{DPS}} = \num{9.63\pm 1.00e7}\,,
		\end{equation}

		\noindent can be used as a good estimate for the reduction factor at nominal conditions, which is very promising with regards to the CPS runtime.
		The runtime of the CPS is limited by the maximal allowed amount of ${N_{\mathrm{CPS,max}} = \SI{0.39}{\milli\bar\litre}}\, {(\estimates \SI{1}{\curie})}$ of accumulated tritium gas. At the nominal tritium gas flow rate of ${Q_{\mathrm{WGTS,d}} = \SI{0.98}{\milli\bar\litre\per\second}}$ from the point of injection in downstream direction, this leads to a maximum operation time before regeneration of:

		\begin{equation}
		t_{\mathrm{CPS, max}} = \frac{N_{\mathrm{CPS,max}}}{Q_{\mathrm{WGTS,d}}} \cdot R_{\mathrm{DPS}} = \SI{445\pm 46}{ days}\,.
		\end{equation}

		\noindent This value surpasses the initial design goal of a CPS regeneration every \SI{60}{ days} by a factor of 7.4. 
		With this rather large safety margin, the measurement interval between subsequent regenerations can be relaxed, allowing for longer neutrino mass runs, and more flexibilty in scheduling of measurements in general.

\section{Summary and conclusion}
	The KATRIN experiment requires a reduction of the tritium flow in the beamline between the point of injection in the WGTS and the spectrometer and detector section by at least \num{14} orders of magnitude. 
	%Otherwise, the background rate of the experiment would increase above the allowed limit by tritium decays inside the spectrometer section and so deteriorate the ultimate sensitivity for the neutrino mass determination. 
	Otherwise, the additional background rate would worsen the ultimate sensitivity for the neutrino mass. 
	%The necessary gas flow rate reduction is achieved by means of the differential pumping section (DPS), and the cryogenic pumping section (CPS), located between the windowless gaseous tritium source and the spectrometer section.
	The huge gas flow reduction is achieved by two sequential pumping systems, each reducing the flow by a factor of at least \num{e7}, using turbo-molecular pumps (DPS) and cryosorption on \SI{3}{\kelvin} cold argon frost (CPS), respectively.
	
	%The tritium reduction factor of the DPS has to be at least \num{e7} to reduce the gas flow rate into the CPS to such a level, that the necessary regeneration of the CPS is only due after more than 60 measurement days. A sound knowledge of the real reduction factor value for high purity tritium is mandatory for the operation of the KATRIN experiment. 
	For the initial design layout, radiation safety considerations required a regeneration of the cryogenic pumping section after no more than 60 days.
	A sound knowledge of the actual reduction factor of the DPS allows for a more accurate estimate of the time interval between regenerations, helping to optimize the time available for neutrino mass measurements.

	%For this reason, extensive gas flow simulations were performed successfully, taking into account the different flow regimes, i.e. transitional and molecular. These simulations proved that the KATRIN pumping concept for tritium reduction is sufficient.
	Therefore, extensive gas flow simulations were performed, taking into account the different flow regimes along the beamline, from laminar and transitional flow to molecular flow.
	The simulation of the DPS resulted in a reduction factor of \num[parse-numbers=false]{8^{+17}_{-6}\times 10^{7}}, well above the minimum requirement.

	To validate the simulations, the tritium reduction factor of the differential pumping sections was measured for the first time in 2019 for different flow rates, with a tritium purity well above \SI{97}{\percent}. 
	
	The measured value for a tritium column density of \SI{4.2e21}{\per\meter\squared} in the beamtube of the WGTS is  $R_{\mathrm{DPS}} = \SI{9.63\pm1.00e7}{}$. 
	%This value is well above the required \num{e7} and can be used as a good estimate for the reduction factor at nominal conditions (column density of \SI{5e21}{\per\meter\squared}), which is very promising with regards to the CPS runtime.
	This reduction factor, measured at \SI{84}{\percent} of the nominal column density, as used during the most recent neutrino runs, is well above the minimum requirement of \num{e7} and is in good agreement with the simulated value.
	
	%In conclusion, the good performance of the differential pumping sections allows the long-term operation of the cryogenic pumping section as intended, enabling the KATRIN experiment to accumulate the necessary amount of measurement runs for its scientific goals.
	In conclusion, the good performance of the final design of the differential pumping section of the KATRIN experiment could be demonstrated both by simulation and measurement for the first time. 
	%It exceeds the design specifications, allowing for longer measurements of the neutrino mass.
	This performance allows the long-term operation of the cryogenic pumping section as intended, enabling the KATRIN experiment to accumulate the necessary amount of measurement runs for its scientific goals.

\section{Acknowledgments}
	
	We acknowledge the support of the Helmholtz Association (HGF), the German Ministry for Education and Research BMBF (05A17VK2), DFG graduate school KSETA (GSC 1085), and the Research Training Group (GRK1694).
	
	We would like to thank Jochen Bonn, Burkhardt Freudiger and Oleg Kazachenko for their ideas and valuable contributions on concepts and design of the experiment and Woosik Gil for his support with the magnet systems and questions related to magnetic fields.
	We would furthermore like to thank Tobias Falke, David Hillesheimer, Joschua Kopheiss, Saskia Sch\"afer, and Tobias Weber for their help in operation of the TLK infrastructure as well as laboratory analyses.
	Last but not least we would like to thank Valery Popov and J\"org Brandt for their contributions to the design and construction of the beam line setup as well as Norbert Kernert for his valuable contributions for the setting up and commissioning of the vacuum components and the tritium loops.

\bibliography{mybibfile_5}

\end{document}